\documentclass[conference]{IEEEtran}
\IEEEoverridecommandlockouts

\usepackage{cite}
\usepackage{amsmath,amssymb,amsfonts}
\usepackage{algorithmic}
\usepackage{graphicx}
\usepackage{textcomp}
\usepackage{xcolor}
\usepackage{xurl}
\usepackage{booktabs}
\usepackage{array}
\usepackage{multirow}
\usepackage{geometry}
\usepackage{makecell}
\usepackage{booktabs}
\usepackage{xcolor}
\usepackage{siunitx}
\usepackage{caption}
\usepackage{hyperref}
\usepackage{times} 
\newcolumntype{C}[1]{>{\centering\arraybackslash}p{#1}}
\newlength{\myarraystretch}
\newlength{\mytabcolsep}
\def\BibTeX{{\rm B\kern-.05em{\sc i\kern-.025em b}\kern-.08em
    T\kern-.1667em\lower.7ex\hbox{E}\kern-.125emX}}

\IEEEpubid{\begin{minipage}{\textwidth}\ \\[60pt]
© 2025 IEEE. Personal use of this material is permitted. Permission from IEEE must be obtained for all other uses, in any current or future media, including reprinting/republishing this material for advertising or promotional purposes, creating new collective works, for resale or redistribution to servers or lists, or reuse of any copyrighted component of this work in other works.
\end{minipage}}

\begin{document}

\title{Hybrid Deep Learning Model for Multiple Cache Side Channel Attacks Detection: A Comparative Analysis\\
}

\author{
\IEEEauthorblockN{Tejal Joshi}
\IEEEauthorblockA{\textit{Dept. of Computer Science \& Engineering} \\
\textit{School of Computational Sciences}\\
\textit{COEP Technological University (COEP TECH)}\\
Pune, India \\
joshits21.comp@coeptech.ac.in}
\and
\IEEEauthorblockN{Aarya Kawalay}
\IEEEauthorblockA{\textit{Dept. of Computer Science \& Engineering } \\
\textit{School of Computational Sciences}\\
\textit{COEP Technological University (COEP TECH)}\\
Pune, India \\
aaryark21.comp@coeptech.ac.in}
\and
\IEEEauthorblockN{Anvi Jamkhande}
\IEEEauthorblockA{\textit{Dept. of Computer Science \& Engineering } \\
\textit{School of Computational Sciences}\\
\textit{COEP Technological University (COEP TECH)}\\
Pune, India \\
jamkhandeaa21.comp@coeptech.ac.in}
\and

\IEEEauthorblockN{Amit Joshi}
\textit{Senior Member, IEEE}\\
\IEEEauthorblockA{\textit{Dept. of Computer Science \& Engineering} \\
\textit{School of Computational Sciences}\\
\textit{COEP Technological University (COEP TECH)}\\
Pune, India \\
adj.comp@coeptech.ac.in}

}

\maketitle


\begin{abstract}

Cache side-channel attacks have emerged as a sophisticated and persistent threat, capable of extracting sensitive user information by exploiting vulnerabilities in modern processors. These attacks leverage the inherent weaknesses in shared computational resources, especially the last-level cache. They infer patterns in data access and execution flows, often bypassing traditional security defenses. These are particularly dangerous because they require no physical access to the victim’s device, making remote attacks feasible. This study focuses on a specific class of these threats— fingerprinting attacks —where an adversary can use cache-side channels to monitor and analyze the behavior of co-located processes, potentially revealing confidential information such as encryption keys or user activity patterns. A comprehensive threat model illustrates how an attacker, sharing computational resources with a target system, can exploit these side-channels. This exploitation allows the attacker to learn patterns in data access, potentially compromising sensitive information. To mitigate such risks, a hybrid deep learning model is proposed to detect cache side-channel attacks. Its performance is compared with five widely-used Deep Learning models like Multi Layer Perceptron, Convolutional Neural Network, Simple Recurrent Neural Network, Long Short Term Memory, and Gated Recurrent Unit, evaluating each model’s resilience to these sophisticated attacks. The experimental results suggest that the hybrid model has achieved a detection rate of up to 99.96\%. The findings demonstrate the limitations of existing models and emphasize the need for enhanced defensive mechanisms, shedding light on future developments for securing sensitive data against evolving side-channel threats.

\end{abstract}


\section{Introduction}
Cache side-channel attacks (CSCA) exploit the shared hardware resources, particularly CPU caches, to extract sensitive information without needing direct access to the target system\cite{b3}\cite{b24}\cite{b35}. By monitoring the cache's behavior, such as timing variations and cache hits or misses, attackers can infer critical data of the users. These attacks pose significant threats across various environments, from cloud computing to personal devices\cite{b36}\cite{b20}. Despite advancements in security, these attacks remain prevalent due to the evolving techniques and complexities of modern systems, making them a persistent concern. Implementing secure architecture is essential to safeguard these environments, as it helps in minimizing vulnerabilities that attackers can exploit through shared hardware resources.

Various kinds of CSCAs, like PRIME+PROBE \& FLUSH+RELOAD, exploit specific aspects of cache behavior. FLUSH+RELOAD involves flushing a shared cache line and monitoring its reloading by another process, revealing access patterns to sensitive data. PRIME+PROBE, in contrast, primes the cache by loading it with the attacker’s information and then probes to detect whether another process has evicted some of this data\cite{b3}\cite{b35}. These techniques, particularly PRIME+PROBE \& FLUSH+RELOAD, are the focus of this study, and their detection relies on Hardware Performance Counters (HPCs) like cache hits, misses, and timing variations\cite{b14}\cite{a3}\cite{b37}. While these counters provide essential indicators, they are often inadequate in identifying sophisticated and evolving attack patterns, necessitating advanced detection methods. This is where deep learning models become invaluable as they can analyze large datasets to detect complex patterns and subtle indicators of an attack that traditional methods might miss\cite{b33}\cite{b13}\cite{b34}. Their proficiency in learning and adapting to new attack vectors makes them essential in modern cybersecurity, especially for detecting CSCAs.

This study makes the following contributions:
\begin{itemize}
    \item Provides a comprehensive comparative analysis of six deep learning models for detecting CSCAs, focusing on both PRIME+PROBE \& FLUSH+RELOAD scenarios.
    \item Proposes the use of a hybrid CNN-LSTM model to improve CSCA detection performance.
    \item Demonstrates the hybrid model consistently outperforming MLP, CNN, RNN, GRU, and LSTM across various attack types, victims, and performance metrics.

\end{itemize} 


Rest of the paper contains Section II that provides an overview of the literature review. The methodology adopted, including the implemented models and the hybrid approach, is detailed in Section III. Section IV presents and analyzes the results of the experiments. The study concludes with key insights and recommendations for future work in Section V.

\section{Literature Review}

\begin{table*}[t]
\caption{\textsc{Recent Research in Cache Side-Channel Attacks' Detection}}
\centering
\setlength{\tabcolsep}{8pt} 
\renewcommand{\arraystretch}{1.3} 
\begin{tabular}{p{4.3cm} p{4.3cm} p{1.8cm} p{2.2cm}}
\toprule
\textbf{Models} & \textbf{Cache Attack(s)} & \textbf{Victim(s)} & \textbf{Perf. Metrics}  \\
\midrule
Machine Learning Models \cite{b18} & PRIME+PROBE, FLUSH+RELOAD, FLUSH+FLUSH & AES & Accuracy, Speed, FP, FN, Overhead \\
MLP \cite{b12} & PRIME+PROBE, FLUSH+RELOAD & AES & Accuracy, FP, FN \\
Change point detection \cite{b16} & PRIME+PROBE, FLUSH+RELOAD, FLUSH+FLUSH & AES, RSA & Detection time, FP Rate  \\
RNN, MLP, LSTM \cite{b11} & PRIME+PROBE, FLUSH+RELOAD, PRIME+ABORT & AES, RSA & Accuracy, FP, FN, Precision, Recall  \\
OneR, MLP, DT, J48, BayesNet \cite{b15} &  PRIME+PROBE, FLUSH+RELOAD & AES, RSA & Accuracy, False Alarm Rate \\
Machine Learning \cite{a5} & FLUSH+RELOAD & Wordpress, Ghost, Chrome & Accuracy, F1-score \\
\bottomrule
\end{tabular}
\vspace{10pt}
\label{table:reference_summary}
\end{table*}

Chiapetta et al. implemented a machine learning technique for detecting FLUSH+RELOAD CSCA in a cross VM environment using \textit{perf} \cite{a3}. Depoix and Altmeyer used machine learning to detect real-time Spectre attacks by collecting performance data using the PAPI tool \cite{a5}. Payer proposed a detection system, HexPADS, to detect PRIME+PROBE \& FLUSH+RELOAD attacks based on system behavior, specifically distinguishing between normal and abnormal activities using a threshold determined by cache-misses \cite{a8}. Briongos et al. proposed a self monitoring tool, CacheShield, to detect FLUSH+FLUSH, PRIME+PROBE \& FLUSH+RELOAD attacks in a cloud-based environment, demonstrating a perfect detection accuracy of 100\% \cite{b16}. Cho et al. introduced a real-time identification framework for CSCA by monitoring in CPU counters using machine learning algorithms, effectively detecting attacks immediately as they occur\cite{a9}.  Su and Zeng proposed a security model that evaluates CSCAs based on vulnerability, cache type, pattern, and range, while also exploring defense strategies \cite{a10}. Alam et al. developed a machine-learning-based identification system for micro-architectural side-channel attacks (SCAs), utilizing performance counter profiling and time-series analysis and achieved a high detection accuracy of 98.7\% \cite{a11}. Wang et al. introduced HybriDG, which is a hybrid model combining Gaussian distribution \& Dynamic Time Warping to detect both recognized and zero-day micro-architectural SCAs in real-time, achieving a detection accuracy of 99.5\% \cite{b26}. Mushtaq et al. proposed WHISPER, a machine learning tool for detecting real-time SCAs like FLUSH+FLUSH, FLUSH+RELOAD, Meltdown, PRIME+PROBE \& Spectre using HPCs and achieved a remarkable detection accuracy of over 99\% \cite{b18}. Wang et al. introduced Hybrid-Shield, a cross-layer solution for detecting and mitigating cache-based side-channel attacks, achieving 100\% detection accuracy with a Decision Tree Classifier \cite{b15}.

Le et al. proposed a real-time detection method for Spectre attacks on RISC-V using HPCs and a Neural Network, achieving over 99\% accuracy with minimal performance impact\cite{b12}. Maheswari and Krishnamurthy brought forth a deep-learning SCA detection method by employing a Deep Residual Capsule Auto-Encoder (DR\_CAE) model for attack classification, achieving high performance with an accuracy of 98.80\%\cite{b28}. Kim et al. introduced a method called FRIME, based on deep-learning for detecting multiple CSCAs, including FLUSH+RELOAD, PRIME+ABORT \& PRIME+PROBE, by leveraging both cache as well as Intel TSX-based hardware events. Their LSTM-based model demonstrated superior performance with a detection accuracy of 98.81\% for FLUSH+RELOAD and 85.33\% for PRIME+PROBE, outperforming their MLP and RNN models \cite{b11}.

Shang et al. proposed a CNN-LSTM hybrid model for processing ultrasonic guided waves in metallic pipelines, achieving damage detection accuracy of 94.8\% as compared to standalone CNN and LSTM models\cite{a12}. Han et al. explored a hybrid CNN-LSTM model for time-series data prediction, addressing the long-term dependency issue inherent in RNNs \cite{a13}. This model showed improved performance in predictive accuracy and power, particularly in dynamic time-series scenarios.

Till now, multiple models have emerged for the detection of CSCAs using machine learning. However, these traditional methods often rely on predefined features and may struggle to adapt to evolving attack patterns. In contrast, deep learning approaches are particularly beneficial because they can autonomously identify and learn intricate patterns from vast datasets without extensive human intervention. Only recently have detection techniques utilizing deep learning models surfaced, addressing the limitations of earlier approaches and offering improved adaptability and accuracy.

Table \ref{table:reference_summary} summarizes the existing findings, detailing models, cache attacks, targeted victims, and performance metrics.

The contribution of this study is driven by the need to improve detection performance and identify the most effective models for CSCA detection. A comparative analysis of various models is crucial to determine the best-performing approaches. Additionally, this study introduces a new hybrid model approach that offers robust protection against these sophisticated threats, comparing it with existing models to solidify its effectiveness.

\section{Proposed Methodology}
This section introduces the proposed hybrid model CNN-LSTM and explores five other deep learning architectures, evaluating their effectiveness against CSCAs. A thorough comparative analysis is conducted to explore the trade-offs and benefits of each model, aiming to identify the optimal configurations that enhance security and detection accuracy without compromising system performance.

\subsection{Implemented Models}

This subsection reviews the deep-learning architectures used in this study, detailing their structure, primary applications, strengths, and limitations. Each model has been implemented for a thorough comparative analysis in the context of CSCA detection \cite{b13}.

\textit{\textbf{Convolutional Neural Networks (CNN):}} CNNs are effective for detecting structured patterns in data. Their ability to capture spatial hierarchies within this data makes them suitable for identifying subtle variations indicative of an attack \cite{b37}. While CNNs can capture local patterns, they don't inherently capture sequential dependencies within the series that may be crucial for detecting attacks patterns  \cite{b40}\cite{b45}.

\textit{\textbf{Recurrent Neural Networks (RNN):}} RNNs are designed to efficiently analyze sequential data, making them applicable for monitoring sequences of cache accesses in SCAs. They excel in learning temporal dependencies, which is crucial for identifying attack patterns over time. Nevertheless, traditional RNNs face challenges like vanishing gradients, which can impair their performance in detecting long-term dependencies \cite{b37}\cite{b34}\cite{b41}.

\textit{\textbf{Long Short-Term Memory (LSTM)}}: LSTM networks identify long-term dependencies in sequential data, making them well-suited for recognizing patterns in time-series data, like in CSCAs. Their ability to retain information over extended sequences allows them to effectively model the temporal dynamics of loads, misses, and instructions, which are crucial for identifying attack behaviors. Although LSTMs require more computational resources compared to simpler models, their strength in handling sequence data justifies their use in this context \cite{b37}\cite{b34}\cite{b42}.

\textit{\textbf{Gated Recurrent Units (GRU):}} GRUs offer a streamlined architecture compared to LSTM networks while retaining the capability to model long-term dependencies in sequential data. Their simplified design enhances computational efficiency, making GRUs a practical choice for detecting CSCAs with reduced resource consumption \cite{b33}. Despite their efficiency, GRUs still require significant computational resources and may exhibit performance trade-offs relative to the more complex LSTM architecture \cite{b43}.

\textit{\textbf{Multi-Layer Perceptrons (MLP):}} MLPs excel at capturing complex non-linear patterns in static data due to their fully connected layers \cite{b33}\cite{b44}. They are effective for tasks where feature interactions are crucial, making them relevant for analyzing extracted features in CSCAs. However, MLPs cannot capture temporal dependencies, limiting their performance on sequential tasks compared to LSTMs, GRUs, and RNNs. Their efficiency and simplicity come at the cost of missing dynamic temporal patterns \cite{b46}.

\subsection{Proposed Hybrid Model} The proposed hybrid architecture integrates CNN and LSTM networks to enhance the detection of CSCAs. This model leverages CNN's capability to extract spatial features from memory access traces, identifying local patterns of loads, misses, and instruction counts. These spatial features are then processed by  the LSTM, which capture temporal dependencies and complex patterns over time. By combining the strength of CNN in spatial analysis with LSTM's ability to model temporal dynamics, this approach aims to improve both accuracy and performance in detecting sophisticated attacks. The hybrid model is anticipated to offer superior detection capabilities compared to models using either CNN or LSTM \cite{b13}\cite{b34}\cite{a12} \cite{a13}.

This model is proposed on observing the diverse set of patterns in the frequency graphs of the data points of the studied attacks. The peaks for the counters of benign and malicious processes are overlapping in some and not in others. On the other hand relatively non-distinct peaks for malicious processes are also observed. These patterns are difficult to detect with a short sequence of inputs, this calls for the use of sequential models. However, distinct peaks between the two are also observed in some cases. Spatial models excel in capturing these instances with their ability to focus on a sliding window of the input. The combination of these abilities can enhance detection.

\begin{figure}[t]
    \centering
    \begin{minipage}[b]{0.48\textwidth}
        \centering
        \includegraphics[width=\textwidth]{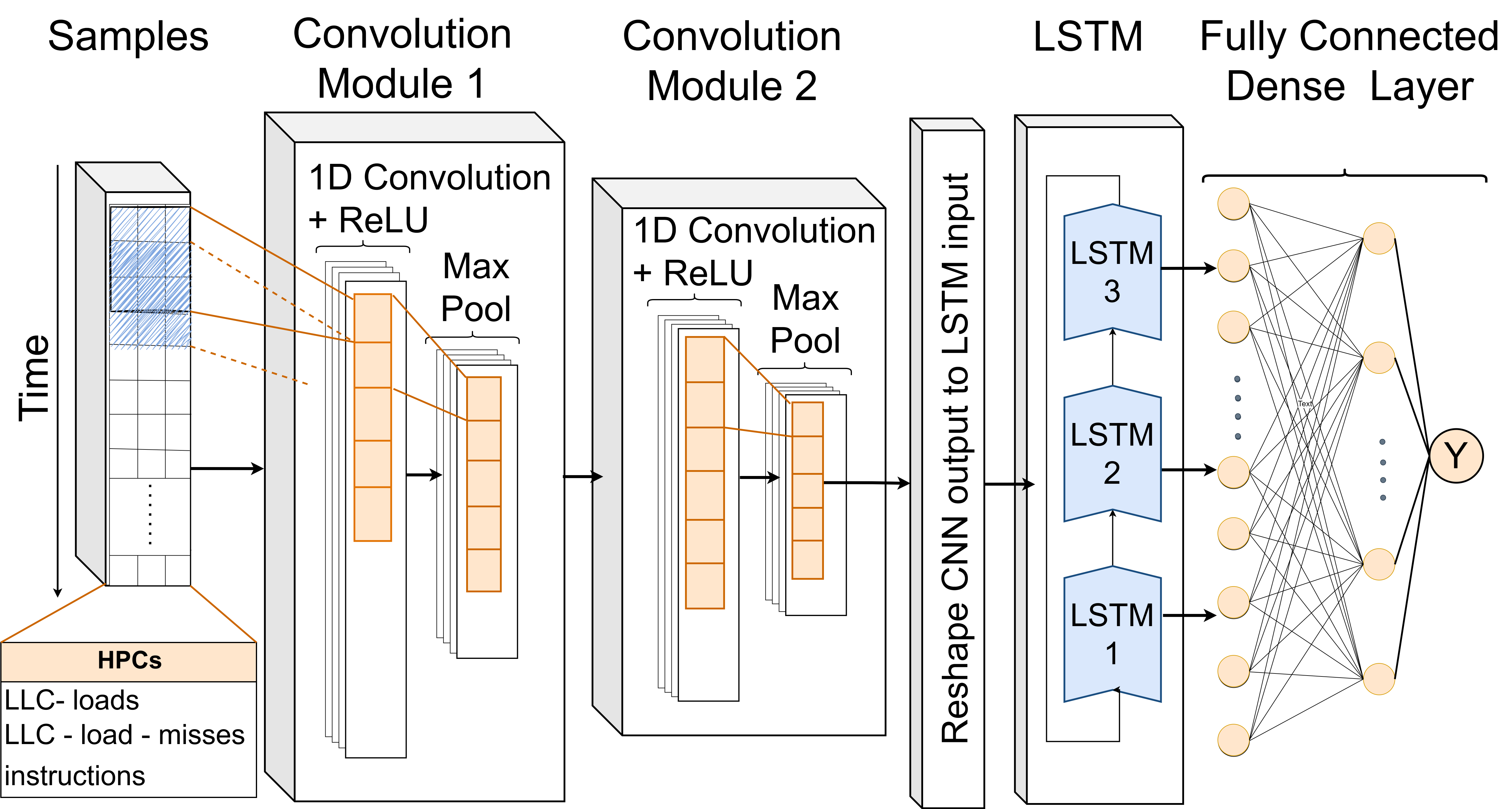}
        \caption{\fontsize{8}{8}\selectfont Block Diagram of Proposed Hybrid Model}
        \label{fig:hybrid}
    \end{minipage}
\end{figure}

Fig. \ref{fig:hybrid} illustrates the flow of data through the hybrid CNN-LSTM model through a block diagram. The model processes input sequences by first passing them through CNN layers, which extract local spatial features using 1D convolution. ReLU activations introduce non-linearity, and max pooling reduces dimensionality while highlighting key features. The CNN output is then fed into LSTM layers, which capture temporal dependencies and long-term patterns within the sequence. Finally, the hidden states from the LSTM are passed through a fully connected layer to produce class predictions based on the combined spatial and temporal features.

This architecture’s main strength is its ability to leverage CNNs for spatial feature detection and LSTMs for temporal pattern recognition, making it well-suited for sequence-based tasks.

\section{Results and Discussion}

This section outlines the implementation details, outcomes, and analysis of the study. 

\subsection{Experimental Setup}
This section explains the environment in which dataset collection, model training and model evaluation is done.

\textit{\textbf{Assumptions \& Considerations:}}
Intel i7-11800H processor with a 16GB RAM and 4MB of Last Level Cache (LLC) is utilized for training and evaluating the models. Processor events are collected on Ubuntu 22.04 using Performance Monitoring Units (PMUs). All the experiments and dataset collection are done on Ubuntu 22.04.

\textit{\textbf{Performance Events:}}
Monitoring specific performance events is crucial for detecting anomalies in CSCAs. 

The \textit{instructions} metric counts the total number of processor instructions executed within a specific time frame. A significant reduction in instruction count coupled with other relevant readings, may indicate malicious activity \cite{a5}.

\textit{LLC-load-misses} occur when data that is not present in the LLC is requested, leading to a cache miss. Spikes in LLC-load-misses can indicate irregular access patterns that may suggest the presence of a CSCA \cite{a8}\cite{a4}.

\textit{LLC-loads} refer to read operations accessing data from the LLC. By tracking LLC-loads, one can assess overall cache usage and identify unusual behavior which when studied with LLC-load-misses, can help in detecting anomalies in cache access patterns.

\subsection{Dataset}
Victim processes running AES and RSA encryption and decryption, are monitored for live data collection\cite{a6}\cite{a7}. Malicious implementations, including PRIME+PROBE \& FLUSH+RELOAD, are also observed. Monitoring is done using the \textit{perf} tool, which leverages PMUs to gather performance data\cite{a2}.

\textit{perf} offers a number of performance events for monitoring purposes. Out of these 2,402 events, three processor events are chosen by considering the runtime behavior of PRIME+PROBE \& FLUSH+RELOAD implementations. The selected events are LLC-load-misses, LLC-loads and instructions.

Performance counters for each process are tracked individually, instead of relying on overall CPU data \cite{a3}. This distinction enables the model to decide whether a process is harmless or potentially malicious. From there, the detection system responds to each process according to the model's predictions.

Overall, 30,000 datapoints are collected and split into a 50-50 distribution, labeled as benign and malicious, respectively. 

Fig. \ref{fig:fr} presents results of FLUSH+RELOAD attack during AES and RSA encryption and decryption, respectively \cite{b39}. 
 
Fig. \ref{fig:pp} displays the results of PRIME+PROBE attack under the same conditions.

\subsection{Evaluation Metrics}\label{sec:Evaluation metrics}
Accuracy, Precision, Recall, FN and FP are some of the widely used metrics for evaluating detection models for CSCAs \cite{b15}\cite{b12}\cite{b11}\cite{b38}. 

\textit{\textbf{Accuracy and Precision:}}
Accuracy measures correct classifications, ensuring overall model reliability. Precision ensures detected CSCAs are true, reducing false alarms.

\textit{\textbf{Recall:}}
Recall is critical for capturing as many CSCAs as possible, minimizing the risk of undetected threats.

\textit{\textbf{False Positives (FP) and False Negatives (FN):}}
FP indicates when normal processes are incorrectly flagged as attacks, while FN shows when actual attacks are missed. Both metrics are crucial for evaluating if the model detects all attacks and avoids unnecessary alerts.

\subsection{Results and Comparison}
\begin{figure}[t]
    \centering
    \begin{minipage}[b]{0.50\textwidth}
        \centering
        \includegraphics[width=\textwidth]{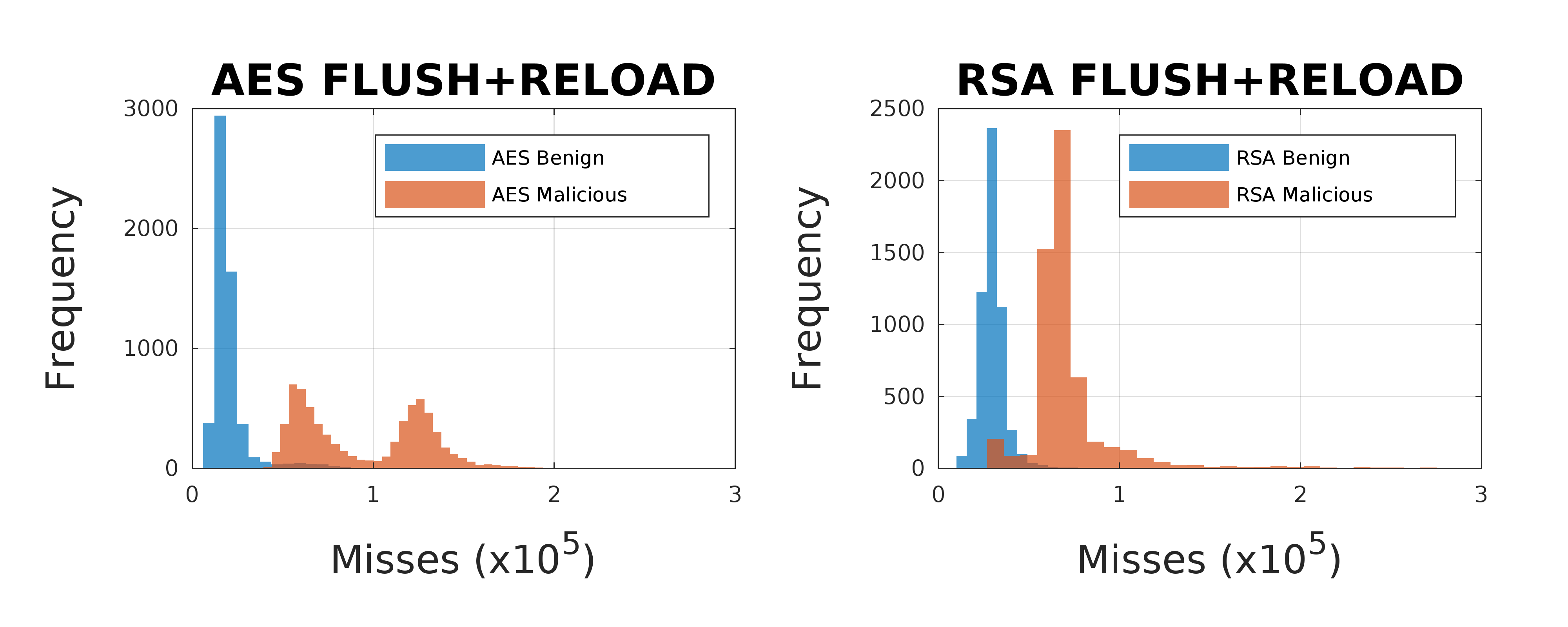}
        \caption{\fontsize{8}{8}\selectfont LLC-load-misses of FLUSH+RELOAD}
        \label{fig:fr}
    \end{minipage}
    \hfill
    \begin{minipage}[b]{0.50\textwidth}
        \centering
        \includegraphics[width=\textwidth]{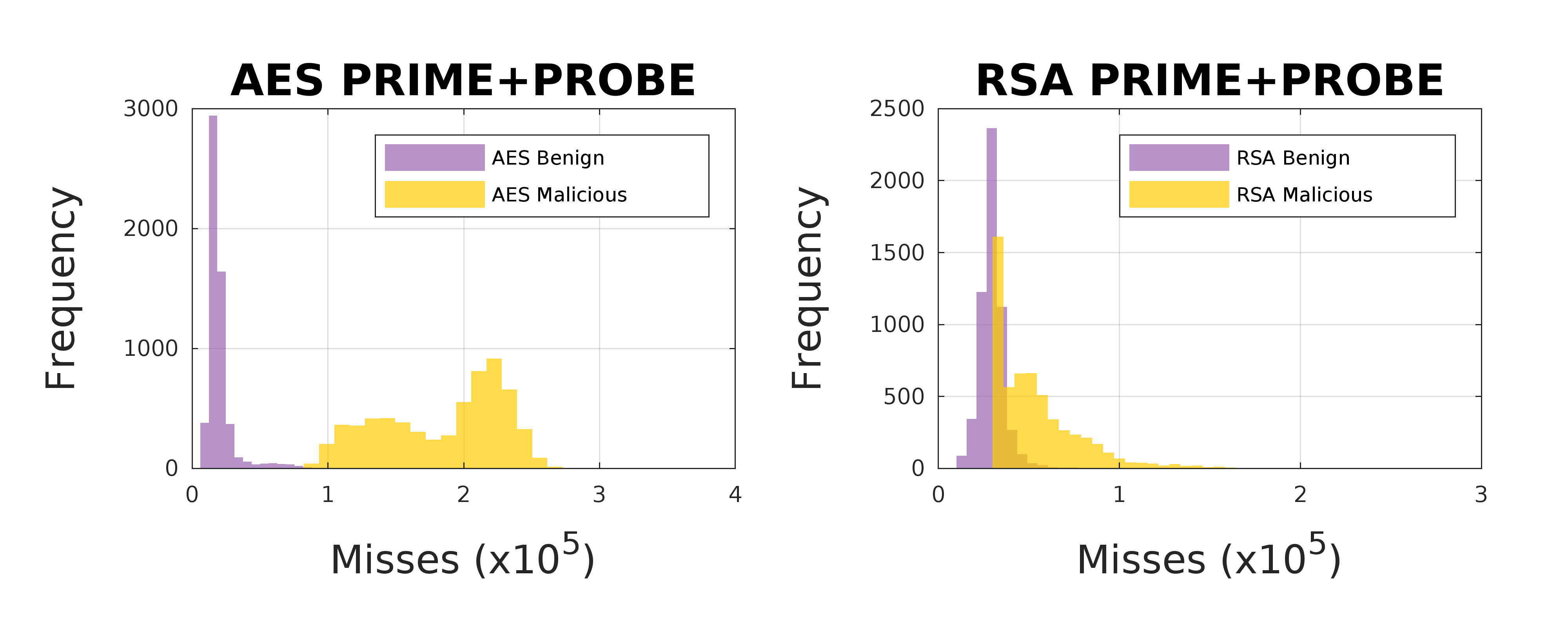}
        \caption{\fontsize{8}{8}\selectfont LLC-load-misses of PRIME+PROBE}
        \label{fig:pp}
    \end{minipage}
    \hfill

  \centering
  \begin{minipage}[b]{0.50\textwidth}
        \centering
        \includegraphics[width=\textwidth]{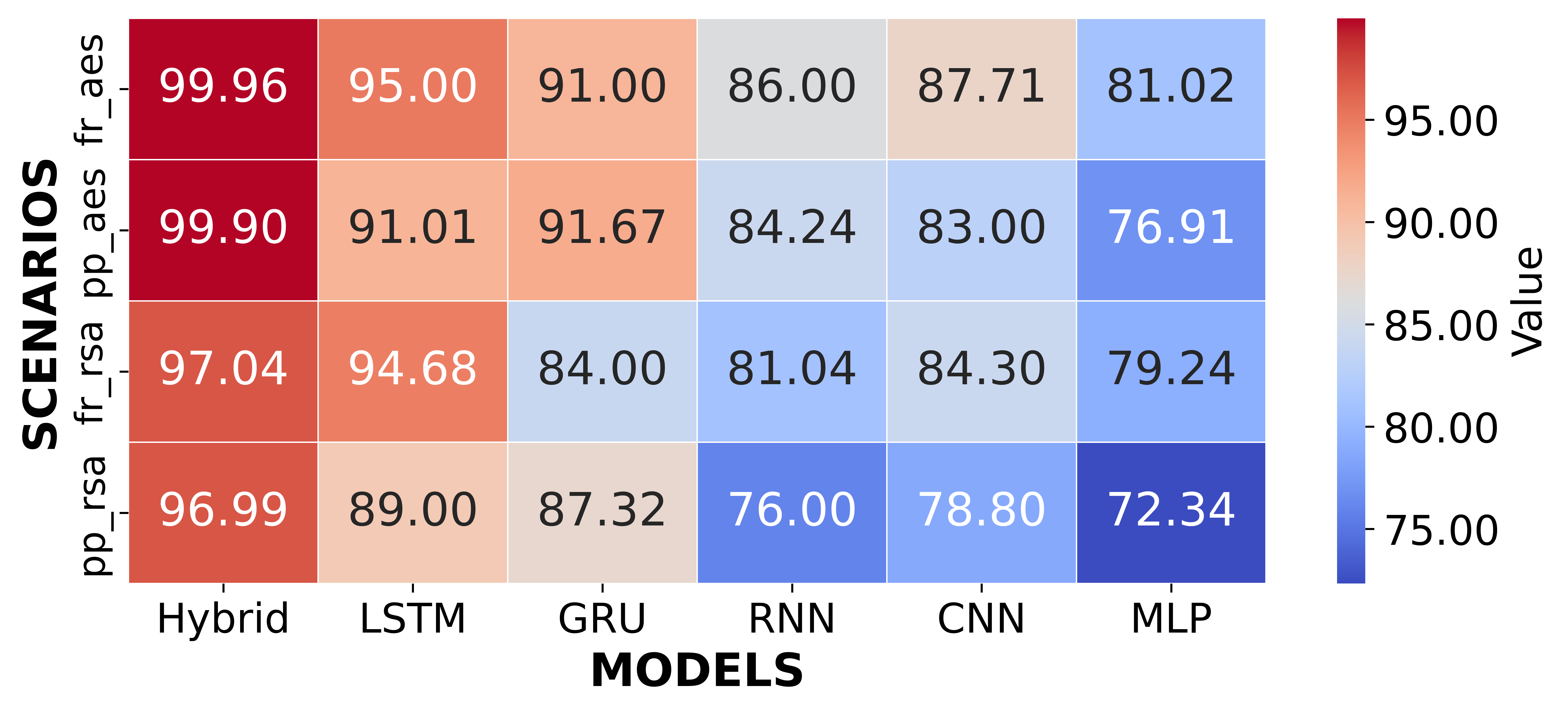}
        \caption{\fontsize{8}{8}\selectfont Heatmap of accuracies of different models}
        \label{fig:heatmap}
    \end{minipage}
\end{figure}

\begin{table*}[!t]
\renewcommand{\arraystretch}{1.25}
\caption{\textsc{Comparison of Different Models Under Diverse Attack Scenarios}}
\label{table:performance_metrics}
\centering
\setlength{\tabcolsep}{10pt} 
\begin{tabular}{|p{1.2cm}|C{2.8cm}|C{1.2cm}|C{1.2cm}|C{1cm}|C{1.2cm}|C{1.2cm}|}
\hline
& \textbf{Model} & \textbf{Accuracy} & \textbf{Precision} & \textbf{Recall} & \textbf{FP} & \textbf{FN} \\
\hline

\multirow{6}{*}{\makecell[tl]{\textbf{FLUSH +}\\ \textbf{RELOAD}\\ \textbf{AES}}} 
& MLP          & 81.02 & 89.51 & 89.51 & 9.49 & 9.49 \\
& CNN          & 87.71 & 92.78 & 94.06 & 6.83 & 5.46 \\
& RNN          & 86.00 & 86.58 & 99.23 & 13.33 & 0.67 \\
& GRU          & 91.00 & 98.91 & 91.92 & 1.00 & 8.00 \\
& LSTM         & 95.00 & 99.99 & 95.01 & 0.01 & 4.99 \\
& {\textcolor{red}{\textbf{Hybrid (CNN-LSTM)}}} & {\textcolor{red}{\textbf{99.96}}} & {\textcolor{red}{\textbf{99.99}}} & {\textcolor{red}{\textbf{99.96}}} & {\textcolor{red}{\textbf{0.00}}} & {\textcolor{red}{\textbf{0.04}}} \\
\hline

\multirow{6}{*}{\makecell[tl]{\textbf{FLUSH +}\\ \textbf{RELOAD}\\ \textbf{RSA}}} 
& MLP          & 79.24 & 88.42 & 88.42 & 10.38 & 10.38 \\
& CNN          & 84.30 & 90.13 & 92.88 & 9.24 & 6.46 \\
& RNN          & 81.04 & 81.64 & 99.11 & 18.23 & 0.73 \\
& GRU          & 84.00 & 97.90 & 85.54 & 1.80 & 14.20 \\
& LSTM         & 91.01 & 94.51 & 96.09 & 5.29 & 3.70 \\
& {\textcolor{red}{\textbf{Hybrid (CNN-LSTM)}}} & {\textcolor{red}{\textbf{99.90}}} & {\textcolor{red}{\textbf{99.95}}} & {\textcolor{red}{\textbf{99.95}}} & {\textcolor{red}{\textbf{0.05}}} & {\textcolor{red}{\textbf{0.05}}} \\
\hline

\multirow{6}{*}{\makecell[tl]{\textbf{PRIME +}\\ \textbf{PROBE}\\ \textbf{AES}}} 
& MLP          & 76.91 & 87.33 & 86.57 & 11.15 & 11.94 \\
& CNN          & 83.00 & 93.61 & 87.99 & 5.67 & 11.33 \\
& RNN          & 84.24 & 89.04 & 93.98 & 10.37 & 5.39 \\
& GRU          & 91.67 & 99.99 & 91.67 & 0.00 & 8.33 \\
& LSTM         & 94.68 & 99.99 & 94.69 & 0.01 & 5.31 \\
& \textcolor{red}{\textbf{Hybrid (CNN-LSTM)}} & \textcolor{red}{\textbf{97.04}} & \textcolor{red}{\textbf{99.99}} & \textcolor{red}{\textbf{97.05}} & \textcolor{red}{\textbf{0.01}} & \textcolor{red}{\textbf{2.95}} \\

\hline

\multirow{6}{*}{\makecell[tl]{\textbf{PRIME +}\\ \textbf{PROBE}\\ \textbf{RSA}}} 
& MLP          & 72.34 & 83.95 & 83.95 & 13.83 & 13.83 \\
& CNN          & 78.80 & 91.51 & 85.01 & 7.31 & 13.89 \\
& RNN          & 76.00 & 81.91 & 91.33 & 16.78 & 7.22 \\
& GRU          & 87.32 & 99.99 & 87.32 & 0.00 & 12.68 \\
& LSTM         & 89.00 & 93.37 & 95.01 & 6.32 & 4.68 \\
& \textcolor{red}{\textbf{Hybrid (CNN-LSTM)}} & \textcolor{red}{\textbf{96.99}} & \textcolor{red}{\textbf{98.61}} & \textcolor{red}{\textbf{98.34}} & \textcolor{red}{\textbf{1.37}} & \textcolor{red}{\textbf{1.64}} \\

\hline
\end{tabular}
\end{table*}

Table \ref{table:performance_metrics} illustrates the results of all the implemented and evaluated models. Each model is tested on its ability to detect PRIME+PROBE \& FLUSH+RELOAD attacks on both AES and RSA as victim programs. The table has four sections to illustrate the results of each type of test viz. FLUSH+RELOAD (AES), FLUSH+RELOAD (RSA), PRIME+PROBE (AES), and PRIME+PROBE (RSA). Each model is evaluated on five metrics which are Accuracy, Precision, Recall, FP \& FN. 

Fig. \ref{fig:heatmap} shows the heatmap of model accuracies across various scenarios and models. 

Two clear patterns are observed within model performance throughout attacks and environments. Firstly, the general pattern observed is that attacks on RSA have a lower detection rate across the board, with accuracy dropping anywhere between 2.00\% to 8.00\% in models detecting the same type of attack. This could be attributed to the fact that RSA has a more irregular access pattern than an AES system \cite{a6}. This results in higher variability in cache miss patterns, making it difficult to distinguish benign activity from malicious activity, i. e. FP in RSA scenarios are much higher \cite{b39}. The same is reflected in the table with gaps up to 6.00\% in the FP values. The attack patterns are muddled by the inconsistent cache behaviour, making it increasingly difficult for spatial models like MLP, CNN; and to a certain extent short sequence temporal models like RNN and GRU to extract meaningful insights from the HPCs collected. The attack pattern distortion created in such RSA environments can only be combated by studying these irregularities on a more longterm scale, hence allowing for effective attack pattern isolation. 

Secondly, this study reveals distinct patterns in the performance of each model, with MLP consistently achieving the lowest accuracy across all test types. However, it exhibits a relatively standout performance in FLUSH+RELOAD scenarios, reaching 81.02\% accuracy, alongside a precision of 89.51\%. There is a clear divide between the accuracy rates it is able to achieve in AES and RSA scenarios, with a gap of roughly 4.00\% in FLUSH+RELOAD and 7.00\% in PRIME+PROBE; one of the highest disparities amongst all models.

CNN performs slightly better with 87.71\% accuracy for FLUSH+RELOAD (AES) but struggles with PRIME+PROBE, especially in the RSA scenario, where it records 78.80\% accuracy. This indicates that despite strong feature recognition, it has limitations in capturing intricate temporal patterns \cite{b45}. It is observed that PRIME+PROBE cache miss patterns occur in bursts or clusters as the attacker probes multiple cache sets, spreading misses periodically over time. In contrast, FLUSH+RELOAD produces localized misses tied to specific cache lines, forming a targeted spatial pattern rather than a temporal pattern.

Hence, spatial models like MLP and CNN, struggle to detect PRIME+PROBE but fare better in detecting FLUSH+RELOAD. CNN overtakes MLP in the former due to its minimal pattern recognition ability being applied when time is taken as one of the spatial dimensions.

RNN performs slightly worse than CNN, with accuracy dropping to 86.00\% for FLUSH+RELOAD (AES) due to its inconsistent spikes. However, it reduces the FN rates by 6\%, exhibiting improved detection of attack patterns. The FP rates increase by roughly 9\%, indicating difficulty in distinguishing attack patterns from benign activity. The vanishing gradient issue likely contributes to these results, as RSA patterns require long-term memory and complex pattern recognition in order to distinguish it from malicious activity.

GRU offers a more balanced performance, achieving roughly 91.00\% accuracy for both PRIME+PROBE \& FLUSH+RELOAD in AES, with up to 98.91\% precision and a roughly equivalent recall of 91.92\%  It significantly improves both FN and FP rates, indicating a  strong ability to differentiate between benign and malicious processes. Notably, it outperforms other models in PRIME+PROBE tests, achieving 87.32\% accuracy for PRIME+PROBE (RSA), where others stagnate at roughly 72.00\%, signalling effective modeling of PRIME+PROBE's temporal attack patterns.

The LSTM model records a 94.68\% accuracy for PRIME+PROBE (AES) and 95.00\% for FLUSH+RELOAD (AES), hinting at its improved performance in temporal pattern recognition over spatial pattern recognition. The FN and FP rates for LSTM remain notabily balanced, showcasing its strength in retaining the afformentioned long-term dependencies in the patterns of PRIME+PROBE and RSA. Unlike GRU, it does not struggle with pattern distinguishing tasks and performs uniformly if not better.

The proposed hybrid CNN-LSTM model outperforms all other models in both PRIME+PROBE \& FLUSH+RELOAD attacks, across both RSA and AES scenarios. It reaches an exceptional 99.96\% accuracy for FLUSH+RELOAD (AES) and 99.00\% for FLUSH+RELOAD (RSA), with the lowest yet FP and FN rates ranging from 0.05\% in AES to a maximum of 3.00\% in RSA. The precision value peaks at 99.99\%, validating the hybrid model's superior ability to capture both spatial and temporal patterns in detecting CSCAs. It hence demonstrates not only pattern recognition but also pattern differentiation, which the previous models do not have the capability to do due to their highly specialized applications allowing only one of these functions at a time. 

This clearly outlines the robustness of the hybrid approach and the value of combining the functional competencies of CNN and LSTM architectures for optimal performance.

The proposed hybrid CNN-LSTM model outperforms the LSTM model implementation by Kim et al. in detecting both PRIME+PROBE \& FLUSH+RELOAD attacks \cite{b11}. This work is preferred for its extensive coverage of attack scenarios, achieving 98.81\% accuracy for FLUSH+RELOAD (RSA) and 85.33\% for PRIME+PROBE (AES). In contrast, the proposed hybrid model achieves 99.01\% and 97.04\% accuracy, respectively. The model also records 99.96\% for FLUSH+RELOAD (AES) and 96.99\% accuracy for PRIME+PROBE (RSA), demonstrating superior performance across multiple CSCAs.

\section{Conclusion and Future Scope}

This study demonstrates the effectiveness of the hybrid CNN-LSTM model in detecting various CSCA scenarios. It combines the spatial feature extraction abilities of CNN with the LSTM’s strength in modeling temporal dynamics. The hybrid approach achieves superior performance in detecting both PRIME+PROBE \& FLUSH+RELOAD attacks, with higher accuracy and precision than all five standalone models. It achieved an overall accuracy of 99.96\%, significantly outperforming the implemented standalone models while also maintaining consistent performance across diverse attack types and environments. This marks a substantial advancement in CSCA detection. Moreover, this study, in comparison to the extensive literature studied in this domain, is amongst the first to provide a robust comparative analysis of these deep-learning architectures in CSCA detection. It is also the first to introduce the use of a proficient hybrid model for detection. It condenses decades of research to extract meaningful insights that could propel further development. 
Another key takeaway from this study is that the long-term pattern recognition capabilities of temporal models like LSTM, GRU, and RNN are enhanced when combined with the localized pattern recognition strengths offered by spatial models such as CNN and MLP.

Further study could focus on expanding the model’s detection abilities to take into account a more diverse set of HPCs. This may allow for the detection of a broader range of attacks. Moreover, future research can focus on optimizing the hybrid architecture to enhance computational efficiency without compromising its performance. Achieving this balance will be crucial in ensuring that the model remains both effective and practical for real-world application scenarios.

\end{document}